\documentclass{aa}  

\usepackage{graphicx}
\usepackage{txfonts}

\begin{document}

\title{A possible wave-optical effect in lensed FRBs}

\author{ Goureesankar Sathyanathan\inst{1, 2},
          Calvin Leung \inst{3},
          Olaf Wucknitz \inst{4}
          \and
          Prasenjit Saha\inst{5},
          }

\institute{Indian Institute of Science Education and Research
              Thiruvananthapuram 69551, Kerala, India\\
              \email{gouree.sath@gmail.com}
        \and
         Department of Physics and Astronomy, University of California, Riverside, CA 92521, USA
         \and
            Department of Astronomy, University of California Berkeley, Berkeley, CA 94720, USA
        \and Max-Planck-Institut für Radioastronomie, Auf dem Hügel 69, 53121 Bonn, Germany
                 \and
             Physik-Institut, University of Zurich, Winterthurerstrasse 190, 8057 Zurich, Switzerland\\
             \email{psaha@physik.uzh.ch}
             }

\abstract
{Fast Radio Bursts (FRBs) are enigmatic extragalactic bursts whose
  properties are still largely unknown, but based on their extremely
  small time duration, they are proposed to have a compact structure,
  making them candidates for wave-optical effects if gravitational
  lensed.  If an FRB is lensed into multiple-images bursts at
  different times by a galaxy or cluster, a likely scenario is that
  only one image is detected, because the others fall outside the
  survey area and time frame.}
{In this work we explore the FRB analog of quasar microlensing, namely
  the collective microlensing by stars in the lensing galaxy, now with
  wave optics included. The eikonal regime is applicable here.}
{We study the voltage (rather than the intensity) in a simple
  simulation consisting of (a)~microlensing stars, and (b)~plasma
  scattering by a turbulent interstellar medium.}
{The auto-correlation of the voltage shows peaks (at order-microsecond separations) corresponding to wave-optical interference between lensed micro-images.  The peaks are frequency dependent if
  plasma-scattering is significant. While qualitative and still in need of more realistic simulations, the results suggest that a strongly-lensed FRB could be identified from a single image.}
{Microlensing could sniff out macro-lensed FRBs.}


\maketitle

\section{Introduction}

Ever since the discovery of an unprecedented signal in an archival
radio time series by D.~J.~Narkevic, then an undergraduate researcher,
and the interpretation of that signal as an exceptionally bright
extragalactic radio transient \citep{2007Sci...318..777L}, Fast Radio
Bursts (FRBs) have steadily increased in interest
\citep{ng2023briefreviewfastradio,Lorimer_2024}.

FRBs last from microseconds to milliseconds, and are characterized by
very high dispersion measures indicating cosmological distances, which
implies that they are 10 to 12 orders of magnitude brighter than
radio sources within the Milky Way. Also, they are very common across the sky,
with a FRB expected once every ten seconds
\citep{Keane_2015,Champion_2016}. While early single-dish telescopes
like Parkes and Arecibo could detect only a few FRBs, the CHIME
telescope \citep{2018ApJ...863...48C} with its much larger field of
view has detected hundreds \citep{2021ApJS..257...59C}.  Observation
of additional signals from FRB121102 led to the understanding that
FRBs can be of a repeating nature
\citep{2014ApJ...790..101S,2016Natur.531..202S}, showing that they are not always cataclysmic events.  While follow-up observations with single-dish telescopes were difficult, CHIME has reported that around 10\% of FRBs detected have a repeating feature \citep{2020Natur.587...54C,CHIMErepeaters},
and a subset of those have periodic activity windows.
Repeating FRBs have wide-ranging properties, peaking at different frequencies,
with the presence of multiple components
\citep[e.g.,][]{Chatterjee_2017,2017ApJ...834L...7T}.
In many cases, precise localization can be carried out, and in these cases, the studies of the intergalactic medium are possible.

The rapid time variation in FRBs indicates small intrinsic sizes.
This would imply that FRB signals are spatially coherent, which would
lead to interference effects in cases of gravitational lensing.
Several authors
\citep{Eichler_2017,2020MNRAS.496..564K,
  2022PhRvD.106d3016K,PhysRevD.106.043017,2023MNRAS.520..247K}
have
considered the possible lensing of FRBs by black holes or other
compact masses.  These works also note a significant complication,
which is frequency dependent plasma lensing by the interstellar medium
in the host galaxy and in the Milky Way.

\cite{Wucknitz_2021} consider multiple images in galaxy and cluster
lensing, and draw attention to the prospect of measuring redshift
drift if time delays from FRB repeaters are observed.  One of the
potential complications noted in that work is the collective
microlensing by stars in the lensing galaxy or cluster.
\cite{2020MNRAS.497.1583L} studied the expected effect of microlensing
on lensing time delays.  A difficulty in finding such multiple-image
systems is that a survey may detect one image from a system, but miss the others because the survey pointing has moved to a different part of the sky.  Since of order one in a thousand high-redshift objects is lensed into multiple images, it is plausible that a strongly-lensed FRB has already been observed without being recognisable as such.

In this work we carry out some simple simulations of wave optics in
collective microlensing by stars.  That is, we combine ideas from the
works mentioned in the previous two paragraphs.  The result suggests
that collective microlensing by stars imprints an observable signature
on a lensed FRB, even without having to detect more than one image.
    
\section{Gravitational lensing with eikonal optics}

The formation of multiple images of a source by an intervening
gravitational field, commonly called strong lensing, occurs in two
very different regimes.  One involves stellar or planetary-scale
masses in or near the Milky Way \cite[see
  e.g.,][]{2024ApJ...976L..19M}.  The other is over cosmological
distances and is the situation more relevant to FRBs.  The theory is
introduced in several places, such as \cite{2022iglp.book.....M} and
\cite{saha2024essentialsstronggravitationallensing} in recent years,
with \cite{2023arXiv230401202L} covering the wave optics part in
detail.  We summarise the essential points below.

\subsection{Images and magnification}

Strong lensing involves a combination of two very different scales.
One is of distances in the combination
\begin{equation}
  {\cal D} \equiv \frac{D_d\: D_s}{D_{ds}}
\end{equation}
Here $D_d$ and $D_s$ are angular-diameter distances from the observer
to the deflector (lens) and the source respectively, and $D_{ds}$ is
the angular-diameter distance from the deflector to the source.
${\cal D}/c$ is of order the total light travel time.
A second scale is
\begin{equation}
  {\cal T} \equiv \frac{4GM}{c^3}
\end{equation}
which is always $\ll {\cal D}/c$.  Image separations are on the scale
of the Einstein radius, given by
\begin{equation}
\label{eq:einstein_radius}
  \theta_E^2 = \frac{c\,\cal T}{\cal D}
\end{equation}
${\cal T}$ itself turns out to be the typical difference in light
travel time between images, known as time delays.  For multiple images
to form $\theta_E$ needs to fit within the angular size of the mass
itself.  This becomes more common for larger distances because
$\theta_E\propto{\cal D}^{-1/2}$ whereas angular size $\propto
D_d^{-1}$.

There are separated subfields of lensing, corresponding to
different scales of $\cal T$ and $\theta_E$.
\begin{itemize}
\item For cluster lenses \citep[for a recent review
  see][]{2024SSRv..220...19N} where $\cal T$ is years to decades, and
  $\theta_E$ reaches the arc-minute scale.
\item For galaxy lenses \citep[see][for a recent
  review]{2024SSRv..220...87S} $\cal T$ is days to months, while
  $\theta_E$ is arc-second scale.
\item Third is the collective lensing by stars within a
  strongly-lensing galaxy or cluster, which splits the individual
  images formed (the so-called macro-images) into many micro-images.
  Here $\cal T$ is of microsecond scale and $\theta_E$ is of
  micro-arcsecond scale.  The phenomenon is of interest for sources
  that are small enough, and has been extensively studied for lensed
  quasars \cite[see
    e.g.,][]{vernardos2023microlensingstronglylensedquasars}.  The
  scale of $\theta_E$ is too small to resolve the images, but changes
  in brightness resulting from transverse motion is observable.  It is
  potentially even more interesting for FRBs, because of the smaller
  source size.
\end{itemize}

Lensing often uses the arrival time surface.  Consider the possible
light travel time for a photon originating at angular position
$\beta$, getting deflected and then arriving from direction $\theta$.
The arrival time can be scaled to a dimensionless form as
\begin{equation}
  t(\theta;\beta) = (1 + z_d) ({\cal D}/c) \, \tau(\theta;\beta)
\end{equation}
and expressed as the sum of geometric and gravitational delays. For a
single point mass the dimensionless arrival time is
\begin{equation} \label{eq:arrivsing}
  \tau(\theta;\beta) = \tfrac12 |\theta - \beta|^2
       - \theta_E^2 \ln |\theta - \theta_0|
\end{equation}
Contributions for more masses can be summed or integrated.  For
collective microlensing by $N$ equal masses we have
\begin{equation} \label{eq:arriv}
  \tau(\theta;\beta) = \tfrac12 |\theta - \beta|^2  - \psi_B(\theta)
       - \theta_E^2\sum_{i=1}^N \mathrm{ln} |\theta - \theta_i|
\end{equation}
with $\psi_B$ representing a background contribution due to all other
masses.  By Fermat's principle, images form where
\begin{equation}
  \nabla \tau (\theta) = 0
\end{equation}
gives the lens equation
\begin{equation}
\label{eq:lens_eq}
  \beta = \theta - \nabla \psi_B(\theta)
  - \theta_E^2 \sum_{i=1}^N \frac{\theta - \theta_i}{|\theta - \theta_i|^2}
\end{equation}

The deflection of light rays by large-scale gravitational lenses like galaxies or galaxy-clusters is described by the macrolensing potential, denoted by $\psi_B(\theta)$ in Eq.~\eqref{eq:lens_eq}. The macrolensing potential governs the overall distortion of light rays originating from background sources. The deflection angle at angular position $\theta$ in the sky is given by the gradient of the potential, $\nabla \psi_B(\theta)$. A Taylor expansion of the deflection angle yields, at leading order, a constant term corresponding to an effective shift in the apparent source position from $\beta$. The next-order term, involving the second derivative of the potential (or the first derivative of the deflection angle), captures the local stretching and shearing of the arrival-time surfaces.  Microlensing simulations \citep[going back to][]{1986ApJ...301..503P} generally assume constant stretching and shearing, neglecting higher-order effects.  For simpler simulations designed for illustrative purposes \citep[e.g.,][]{vernardos2023microlensingstronglylensedquasars} the background potential may be neglected altogether.  Its omission does not qualitatively affect our primary results.

By visualizing the arrival-time surfaces $\tau (\theta;\beta)$ it is
possible to visualize three kinds of images formed at locations where
the surface has zero gradient.  If there is no lens, then there is a
minimum at the point $\theta = \beta$ surrounded by a parabolic
well. A circular lens directly aligned with the source replaces this
parabola with a hill surrounded by a circular valley. However, if this
is not the case, the valley will be asymmetric with a bowl on one
side. The more complex the mass distribution, the more valleys and
hills appear, leading to maxima and minima. Another set of images is
formed by saddle points that also satisfy the zero-gradient condition.

The Hessian of $ \tau (\theta)$ determines how
quickly or slowly we will be moved off-source by a change in
$\theta$.  By defining $(M^{-1})_{ij} \equiv \partial_i \partial_j \tau (\theta)$ as
the inverse magnification matrix, it can explain the lensing caused by
many small and finite sources as
\begin{equation}
   \Delta \theta_i = \sum_j \partial_i\partial_j\tau(\theta) \Delta \beta_{j}
\end{equation}
The scalar magnification $\mu(\theta)$ is given by
\begin{equation}
   1/\mu(\theta) = \det \big( \partial_i\partial_j \tau(\theta) \big)
\end{equation}
Note that the magnification is negative for images at saddle points.
    
\subsection{Wave optics in the eikonal regime}

While geometric optics is a good enough approximation for most
astronomical sources that are large enough so radiation from different
parts of the source reaches the observer incoherently, wave optics
applies to small and far away sources.  This applies to gravitational
waves \citep[e.g.,][]{dai2020searchlensedgravitationalwaves,
  2021MNRAS.508.4869M}.  The
case of FRBs is actually simpler, because of the much higher
frequencies, enabling an eikonal approximation.

An unlensed narrow-band signal of the form
\begin{equation}
        S_{UL} = \sum_\nu \: S(\nu) \: \exp(2 \pi i \nu t)
\end{equation}
results in a lensed signal
\begin{equation} \label{eq:lensedsignal}
         S_{L} = \sum_\nu \: A(\nu) \: S(\nu) \: \exp(2 \pi i \nu t)
\end{equation}
where we have defined
\begin{equation}
        A(\nu) = \sum_k \alpha_k \exp(2\pi i \nu \tau_k)
\end{equation}
\begin{equation} \label{eq:phasecases}
          \alpha_k = |\: \mu_k\: |^{1/2} \times \begin{cases}
            1 & \mathrm{for \; minima}\\
            i & \mathrm{for \; saddle \; points}\\
            -1 & \mathrm{for \; maxima}
        \end{cases}
\end{equation}
and $\tau_k$ corresponds to the $k$-th micro-image.  The lensed signal
from a large number of micro-images is thus straightforward to
calculate.

\subsection{Plasma scattering}

In addition to gravitational lensing, an important obstacle in the
radio domain is the scattering caused by turbulence of the
interstellar medium. This depends on the electron density and amounts to an additional time delay that is proportional to $\nu^{-2}$. This produces deflections similar to those produced by microlensing, forming a large number of images that can interfere with each other.  Due to their extragalactic nature, FRBs can be affected by scattering in the lens galaxy and the intergalactic medium apart from the Milky Way. This would be most serious if the lensing galaxy is gas-rich, but as most lensing galaxies are elliptical, a comparatively low level of plasma scattering is expected \citep{Wucknitz_2021}.

To model plasma scattering we add a term
\begin{equation} \label{eq:turbul}
   (\nu_0/\nu)^2 \; \theta_E^2 \; p(\theta)
\end{equation}
to the scaled time-delay \eqref{eq:arriv}.  Here $p(\theta)$ denotes a
turbulent field normalized as
\begin{equation} \label{eq:turbulnorm}
   \left\langle p^2(\theta) \right\rangle = 1
\end{equation}
The strength is then set by $\nu_0$, which we can understand as the
frequency for which plasma scattering contributes as much as a star
microlensing on its own.  Gravitational lensing dominates in the
$\nu\gg\nu_0$ regime. Plasma scattering becomes less and less important with an increase in $\nu$.

\begin{figure}
\centering
\includegraphics[width=\hsize]{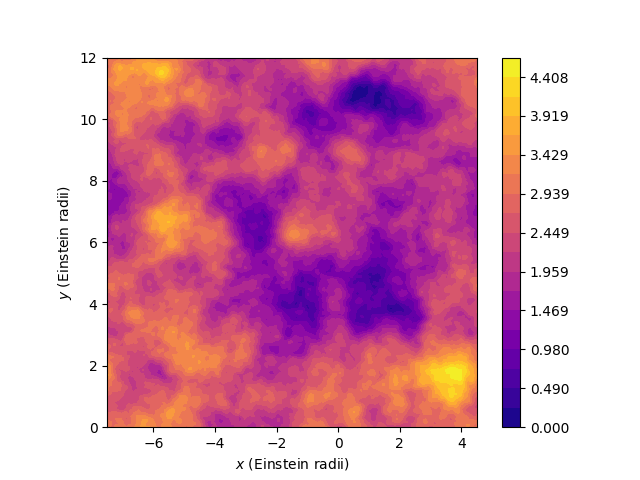}
\caption{Illustration of time delay due to turbulent ISM, normalized
  according to Eq.~\eqref{eq:turbulnorm}.  The spatial scale
  is the same as the microlensing fields in left panels in
  Figs.~\ref{fig:highfreq}--\ref{fig:lowfreq} below.
  \label{fig:turbul}}  
\end{figure}

To produce a realization of $p(\theta)$ we let its Fourier transform
be a power law with small-scale and large-scale cutoffs
\citep[cf.~Eq.~1 in][]{1995ApJ...443..209A}.  The Fourier component
corresponding to
\begin{equation}
   (q_x,q_y,q_z)
\end{equation}
is given an amplitude of
\begin{equation}
\left(q^2 + L_0^{-2}\right)^{-11/6} \exp\left(-\tfrac12 l_0^2\,q^2\right)
\end{equation}
and a random phase.  Here $l_0$ and $L_0$ are the grid-size and
field-size respectively, which act as the small-scale and large-scale
cutoffs.  Since an integral along the line of sight is desired, we put
$q_z=0$.  The real part of $p(\theta)$ is taken, and normalized
according to Eq.~\eqref{eq:turbulnorm}. Fig.~\ref{fig:turbul} shows an example realization of $p(\theta)$.

The grid-size for the simulation is chosen as $1024 \times 1024$, with each point corresponding to one Einstein radius. The physical distance corresponding to the Einstein radius (Eq.~\ref{eq:einstein_radius}) is given by
\begin{equation}
    {\cal D} \, \theta_E = \sqrt{c\,\cal T \, D}
\end{equation}
and is of order milli-parsecs or light-days. The small-scale cutoff $l_0$ is roughly $10^{-2} {\,\cal D\,}\theta_E$ and the large-scale cutoff $L_0$ corresponds to $10 {\,\cal D\,}\theta_E$.  The region of interest where microlensing is assumed to take place lies between these two scales, and the turbulence is expected to be scale-free.

\section{A simple simulation}

We present in this section the results obtained from the effects of
macrolensing, microlensing, and a turbulent interstellar plasma on the FRB signal.

We start with illustrative microlensing simulation from the recent literature \citep[][---see especially their Figs.~4 and 6]{vernardos2023microlensingstronglylensedquasars}.  The simulation has 50 equal point-mass lenses, whose Einstein radii cover $\sim40\%$ of the image plane ($\kappa_*\sim0.4$ in the usual notation).  This assume that the stellar mass is a significant contributor to the total lensing mass, but the stellar mass need not be dominant.  More elaborate simulations have many thousands of particles, and include a smooth tidal field (known as external shear), but the qualitative properties of micro-images are already present in this example.  Of particular interest is a cluster of four bright micro-images, two minima and two saddle points.

To the scaled time delay \eqref{eq:arriv} corresponding to the above lensing system, we add the plasma scattering term~(\ref{eq:turbul}) with adjustable $\nu_0$. The micro-images are found using a simple recursive grid search. A random coherent signal spectrum $S(\nu)$ is generated and this signal is lensed according to Eqs.~(\ref{eq:lensedsignal}--\ref{eq:phasecases}). The lensed signal is then auto-correlated. 
Using Eqs.~(\ref{eq:lensedsignal}--\ref{eq:phasecases}), the auto-correlation turns out to be 
\begin{equation}
    \label{eq:autocorr}
    \begin{aligned}
        C(t) &= \int S_L(t')\, S_L^*(t'-t) \, dt' \\
        &= \sum_\nu \sum_{j,k} \alpha_j \alpha_k^* \,|S(\nu)|^2 \, \exp(2 \pi i \nu (t + \tau_j - \tau_k)) 
\end{aligned}
\end{equation}
$C(t)$ clearly has peaks when $t = \tau_k  - \tau_j$.  For a theoretical understanding, we distinguish between the real and imaginary parts of the auto-correlation, even though they may not be distinguishable in practice.

\subsection{Lensing-dominated regime}
\begin{figure*}
\centering
\setlength{\tabcolsep}{1mm}  
\renewcommand{\arraystretch}{1}   
\begin{tabular}{cc} 
   \includegraphics[width=0.45\textwidth]{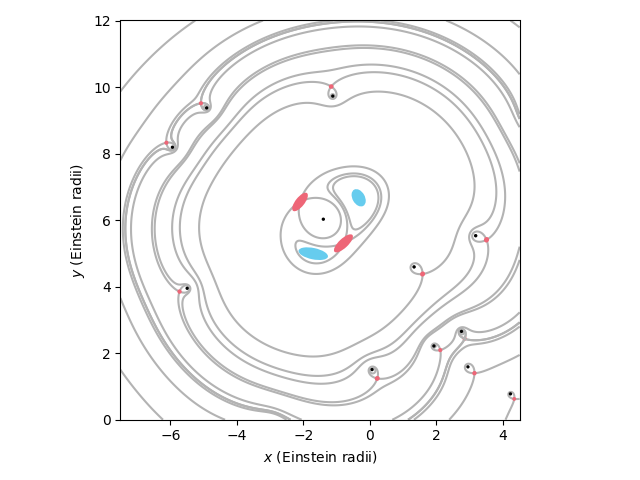} &
   \includegraphics[width=0.45\textwidth]{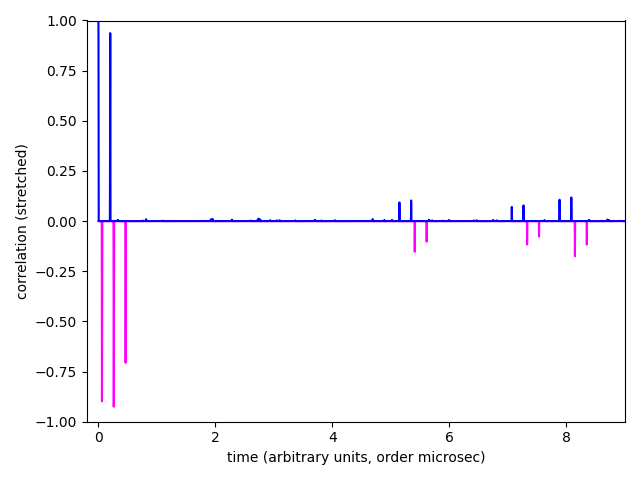} \\
\end{tabular}
\caption{\textit{Left.}~Micro-images in the lensing-dominated regime.
  Black dots mark stellar masses.  Gray curves show arrival time
  contours.  Ellipses are micro-images (blue for minima, red for
  saddle points). \textit{Right.} Auto-correlation signals: blue for
  the real part, magenta for the imaginary part. A tanh activation
  function has been applied to reduce the contrast between the
  lines. \label{fig:highfreq}}
\end{figure*}

Fig.~\ref{fig:highfreq} illustrates this regime.
The left panel in Fig.~\ref{fig:highfreq} is equivalent to Fig.~6 (up to some cosmetics) in \cite{vernardos2023microlensingstronglylensedquasars}. 
The micro-images present are as follows.
\begin{itemize}
\item Two bright minima with time delays (say) $\tau_0, \tau_1$.
\item Two bright saddle points with $\tau_2 , \tau_3$.
\item Faint saddle points with $\tau_4 , \tau_5, \tau_6 \ldots$
\end{itemize}
An important property of collective microlensing by stars \citep[see
  e.g.,][]{1984JApA....5..235N,2011MNRAS.411.1671S} is that the
observable signal is completely dominated by a few bright
micro-images, while the faint micro-images, though much more numerous,
have a negligible contribution to the light.  In this example, the two
bright minima and two bright saddle points all have $|\mu_k|\sim1$
with the remaining images being orders of magnitude fainter.  Thus
only the first four micro-images would be observationally relevant.
For a theoretical understanding, however, we will continue to consider
the first few faint micro-images as well.

The right panel in Fig.~\ref{fig:highfreq} shows the auto-correlation (Eq.~\ref{eq:autocorr}) of the lensed signal in the left panel of the same figure. Each line in the
auto-correlation (after the line at zero delay) is associated with a
pair of micro-images.  The strong lines are associated with pairs of
bright micro-images. Blue and magenta indicate the real and imaginary parts.

\begin{itemize}
\item At $\tau_1-\tau_2$ between two opposite-parity images, and
  therefore imaginary.
\item At $\tau_3-\tau_2$ and $\tau_1-\tau_0$ which happen to be almost
  equal, between like-parity images, and therefore real.
\item At $\tau_2-\tau_0$ and $\tau_3-\tau_1$ which again happen to be
  nearly equal between like-parity images.
\item At $\tau_3-\tau_0$ again between opposite parity images.
\end{itemize}
Then, there is a repeating pattern of four lines, as a faint
micro-image is paired with the four bright micro-images
\begin{itemize}
\item at $\tau_k-\tau_0$ and $\tau_k-\tau_1$ between opposite-parity
  images, and
\item at $\tau_k-\tau_2$ and $\tau_k-\tau_3$ between like-parity
  images
\end{itemize}
for $k=4,5,6\ldots$ which continues beyond the figure.  These are only
of theoretical interest, as there is no prospect of observing them.
There are also some even fainter lines
\begin{itemize}
\item at $\tau_k-\tau_j$
\end{itemize}
for $k,j>3$ corresponding to pairs of faint micro-saddles.

The magenta lines in Fig.~\ref{fig:highfreq} all happen to be negative.  The reason for this can be seen by referring back to Eq.~(\ref{eq:autocorr}) for the auto-correlation.  Since the figure shows positive $t$, it implies $\tau_k>\tau_j$ at the peaks.
The magenta lines come from cases where $\tau_j$ is a minimum and $\tau_k$ is a saddle point.  Since all the saddle-points arrive later than all the minima, $\tau_k$ always corresponds to a saddle point, and the associated $\alpha_k^*$ factor makes the imaginary part negative.  If there would be any saddle points arriving earlier than a minimum, a positive magenta line would appear.




\subsection{The effect of plasma scattering}

We now show the effect of adding the plasma-scattering term
(\ref{eq:turbul}).  Fig.~\ref{fig:highfreq} discussed above actually
already had a small plasma-scattering contribution, corresponding to
$\nu=20\,\nu_0$.  In the next two figures, we gradually reduce $\nu$.  In these figures we also zoom in to highlight the bright micro-images and the auto-correlation peaks between them.

\begin{figure*} 
\centering
\setlength{\tabcolsep}{1mm}  
\renewcommand{\arraystretch}{1}   
\begin{tabular}{cc} 
    \includegraphics[width=0.45\textwidth]{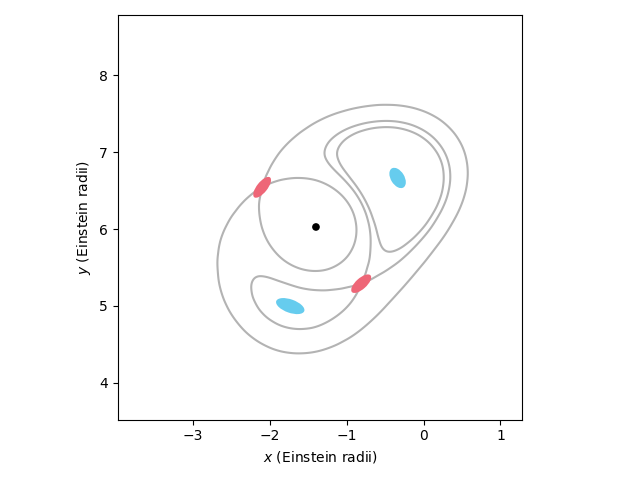} &
    \includegraphics[width=0.45\textwidth]{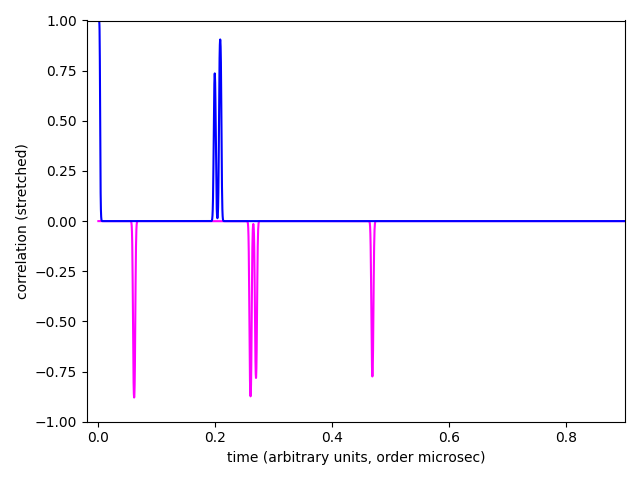} \\
    \includegraphics[width=0.45\textwidth]{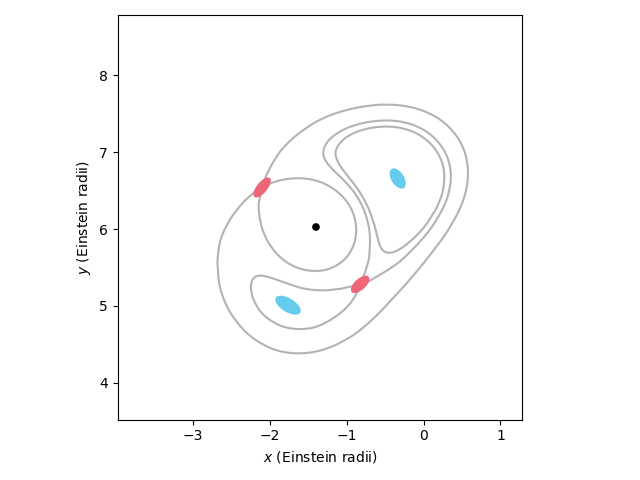} &
    \includegraphics[width=0.45\textwidth]{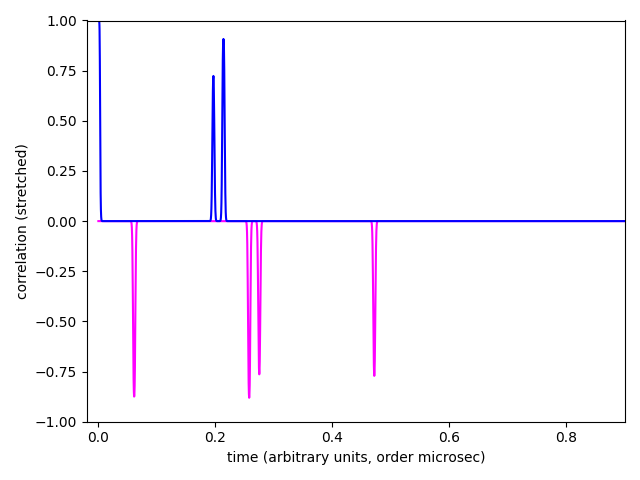} \\
    \includegraphics[width=0.45\textwidth]{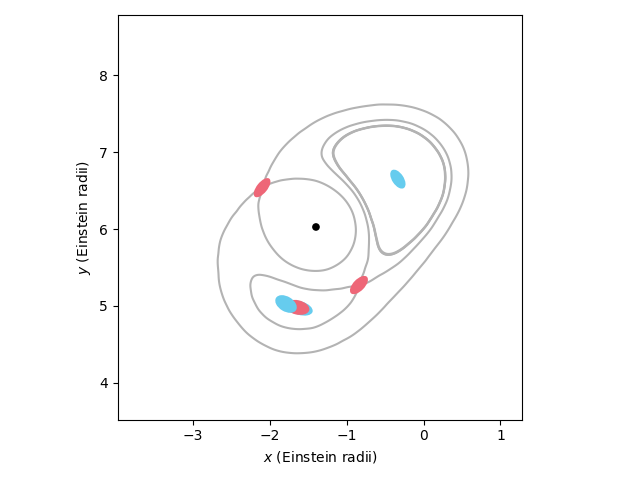} &
    \includegraphics[width=0.45\textwidth]{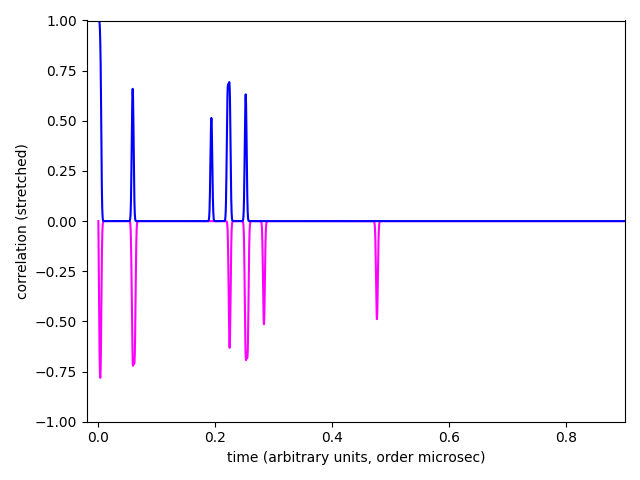} \\
\end{tabular}
\caption{Arrival time contours (left) and auto-correlation signals
  (right) for intermediate-frequency signals at $\nu/\nu_0$ of 13 (top
  row), 10 (middle row), 8 (bottom row) \label{fig:midfreq}}
\end{figure*}

In Fig.~\ref{fig:midfreq}, we see that down to $\nu=10\,\nu_0$ there is only a slight shift in the auto-correlation peaks; that is, plasma scattering is not
significant.  When we reduce to $\nu=8\,\nu_0$ however, a qualitative
change appears.  The bright micro-minimum initially at $\tau_2$ splits
into three bright micro-images (two minima and a saddle point).
Accordingly, more lines appear in the auto-correlation.

\begin{figure*} 
    \centering
    \setlength{\tabcolsep}{1mm}  
    \renewcommand{\arraystretch}{1}   
    \begin{tabular}{cc} 
        \includegraphics[width=0.45\textwidth]{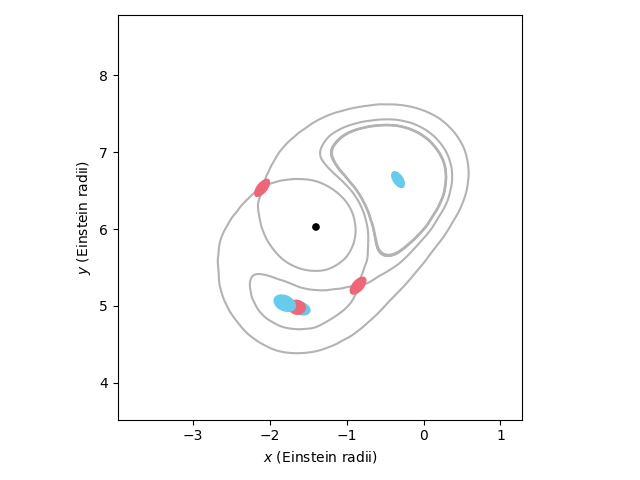} &
        \includegraphics[width=0.45\textwidth]{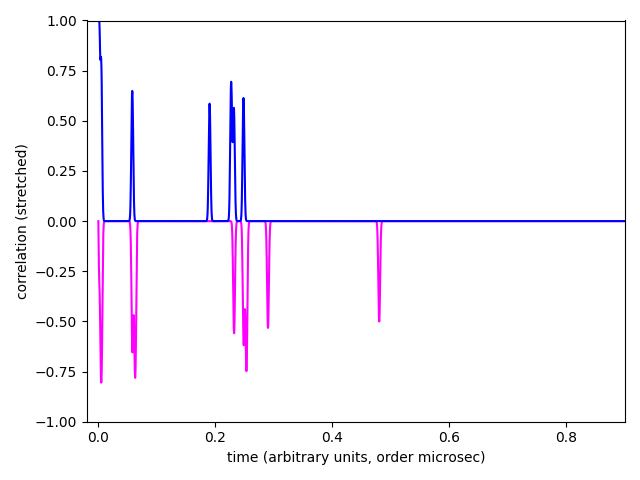} \\
        \includegraphics[width=0.45\textwidth]{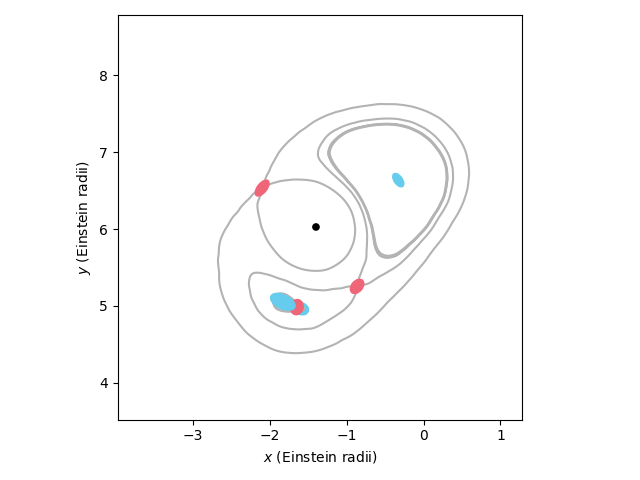} &
        \includegraphics[width=0.45\textwidth]{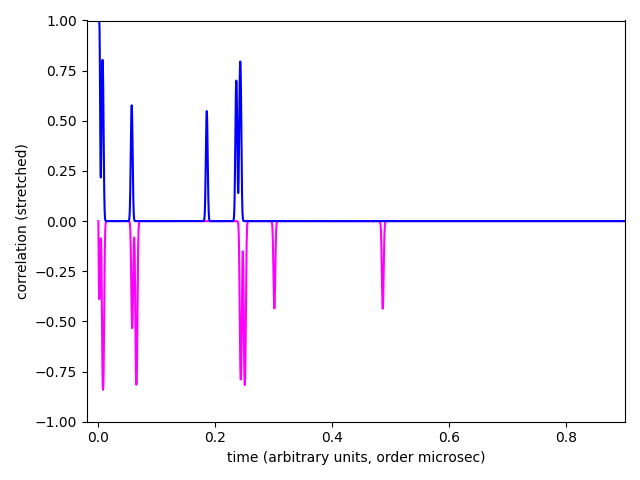} \\
        \includegraphics[width=0.45\textwidth]{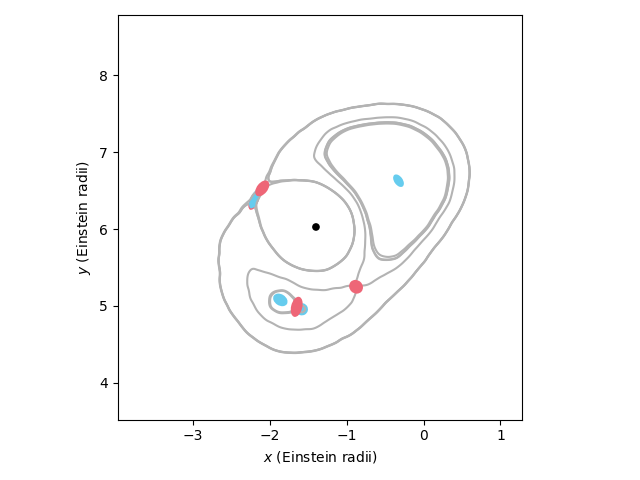} &
        \includegraphics[width=0.45\textwidth]{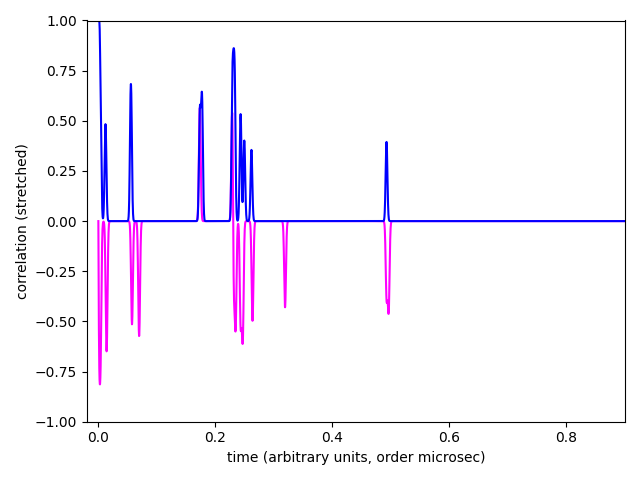} \\
    \end{tabular}
\caption{Arrival time contours (left) and auto-correlation signals
  (right) for low-frequency signals, $\nu/\nu_0$ being 7 (top row), 6
  (middle row), 5 (bottom row). The slight wiggliness of the curves in
  the bottom left panel are due to small-scale fluctuations from
  plasma scattering. \label{fig:lowfreq}}
\end{figure*}

In Fig.~\ref{fig:lowfreq}, the frequency is lowered in steps to
$5\,\nu_0$.  The saddle point initially at $\tau_3$ now splits into
three images, while the triplet near $\tau_2$ that formed at
$\nu=8\,\nu_0$ separates out a little.  Here we have the first example of a minimum (near $\tau_3$) that is later than a saddle point (near $\tau_2$), leading to a positive imaginary part.

We thus see the familiar theme, that the observable is frequency
independent if gravitational lensing dominates, and frequency
dependence if interaction with the medium is important.

\section{Observational aspects}

Details of the observations and data processing go beyond the scope of this theoretically motivated manuscript, but we need to understand the chance of an actual detection of microlensing signatures.

FRBs are basically found as peaks in time series of the detected power over long observations. In reality, the data are channelised to form dynamic (time-resolved) spectra, to be able to correct for dispersion from the interstellar medium, and to distinguish astrophysical bursts from artificial interference. For the microlensing analysis, the original voltage signal $S(t)$, or equivalently $S(\nu)$, is needed for the duration of the burst, to form the autocorrelation according to Eq.~\eqref{eq:autocorr}. Such data are now recorded on a regular basis by CHIME and ASKAP, but also by telescopes like Effelsberg, which are used for follow-up observations of repeating bursts.

Because the central peak at $C(0)$ corresponds to the integrated detected power, we know it to be well above the noise level. Microlensing-induced peaks at $\tau_k-\tau_j$ (for $k>j\geq0$) will be proportional to $|\alpha_j\alpha_k|$ or the geometric mean of the intensities of the corresponding micro-images. This can be compared to the central peak, which is effectively proportional to $\sum_j |\alpha_j|^2$ (the sum of all intensities), after averaging over frequency.

For the image configuration in Fig.~\ref{fig:highfreq}, the ratio is 0.29 for the highest microlensing peak compared to $C(0)$, which is detectable if the initial FRB detection has a sufficient signal-to-noise ratio. The actual detection analysis requires a thorough statistical analysis of false-detection rates, taking into account the number of trial lags.

The frequency (here $\nu_0$) at which plasma scattering and lensing
become comparable is unknown.  If $\nu_0$ is lower than the GHz scale
of FRB observations, the wave-optical effect discussed here will be
observable.  Otherwise the effect will be washed out by plasma
scattering.  The overall delay due to plasma scattering will be
comparable to the observed dispersion (which is of order a second).
This is vastly larger than the microsecond scale delays due to
individual stars, but tiny compared to the overall time delay of to
all the lensing mass (time delays in galaxy lensing are months).  What
is relevant is whether the plasma-scattering term varies on the
microsecond scale over the transverse scale of microlensing (which is
micro-arcseconds or picoradians).  Equivalently, the question is
whether the small-scale cutoff of the turbulent plasma is smaller than
the microlensing scale.

The scattering delays can depend on the galactic latitude, with low galactic latitudes showing stronger effects in general \citep{Wucknitz_2021}. For latitudes above $20^ \circ$, a scattering delay at 1.4GHz is always less than 1 microsecond. 
Additional constraints on small-scale plasma fluctuations come from secondary-spectrum measurements \citep[Fig. 4]{2021MNRAS.500.1114S}. Apart from a special case at 1 millisecond, most scattering features have delays of the order of a few 100 microseconds at 330MHz. As typical delays scale as $\sim \nu^{-4.4}$, even the strongest 1 millisecond feature would be $\sim 1.7$ microseconds at 1.4GHz - a case of strong scattering. Characteristic scattering angles at about 330MHz are also found to be $\sim 30$mas, which scale with frequency as $\nu^{-2.2}$, implying angular broadening of order $\sim 1$mas at 1.4GHz \citep[Fig. 9]{2021MNRAS.500.1114S}. So, it turns out microseconds and mas are appropriate order-of-magnitude estimates at 1.4GHz. In the lensing galaxy however, scattering can be substantially stronger as the delays scale with distance. In such environments, plasma phase fluctuations may approach or exceed microlensing scales and could potentially suppress wave-optics features.

\section{Conclusions}

Based on the above simulation, we make the following interpretation. If a lensed FRB is smaller than the Fresnel scale --- which is $\left({\cal D}\,c/\nu\right)^{1/2} \sim 10\,$au for GHz observations --- and moreover is not smeared too much by the contents of the interstellar or intergalactic medium --- making it a coherent source --- then a single macro-image could be used for recognizing the lens.  The well-known fine time structure in FRBs indicates that they are sufficiently small to be coherent.  Since most strongly lensing galaxies are elliptical, lensed FRBs would be exposed to the low end of plasma scattering.

It would be interesting to examine auto-correlated voltage time series
at different frequencies \citep[as simulated
  in][Figs.~11--12]{2024PhRvD.110l3027K} for a sample of FRBs observed
in different parts of the sky and thus through different regions of
the Milky Way.  While not expected to yield any lensed specimens, even a small sample would give an indication of the importance of plasma scattering through the frequency dependence of the auto-correlation.  If there are frequency-dependent peaks in the
auto-correlation at all frequencies, it would not be possible to separate the contributions of gravitational lensing and plasma scattering.  If auto-correlation peaks are absent at
some (high) frequencies, the outlook for finding lenses by this method
would be positive.

\bibliographystyle{aa} 
\bibliography{refs}

\end{document}